\documentclass{aa}
\usepackage{graphicx}
\usepackage{longtable}
\usepackage{multirow}
\usepackage{txfonts}
\newcommand{\kms}{km\,s$^{-1}$}       %km/s

\newcommand{\um}{$\mu$m}                                 %micron
\newcommand{\hcop}{HCO$^+$($1\rightarrow 0$)}

\newcommand{\sio}{SiO(v=0, $2\rightarrow 1$)}
\newcommand{\sii}{[Si{\sc ii}]34.8\um}
\newcommand{\neii}{[Ne{\sc ii}]12.8\um}
\newcommand{\feii}{[Fe{\sc ii}]26\um}
\newcommand{\lsun}{L$_{\odot}$}               %solar and terr.units
\newcommand{\msun}{M$_{\odot}$}

\newcommand{\rsun}{R$_{\odot}$}
\newcommand{\mdot}{\.{M}}

\newcommand{\msunyr}{M$_{\odot}$\,yr$^{-1}$}

\newcommand{\lsim}{\;\lower.6ex\hbox{$\sim$}\kern-7.75pt\raise.65ex\hbox $<$\;}
\newcommand{\gsim}{\;\lower.6ex\hbox{$\sim$}\kern-7.75pt\raise.65ex\hbox $>$\;}
\newcommand{\amin}{$^{\prime}$}                   %arcus and coordinates
\newcommand{\asec}{$^{\prime \prime}$}
\newcommand{\adeg}{$^{\circ}$}
\newcommand{\pam}{.\hskip-2pt$^{\prime}$}
\newcommand{\pas}{.\hskip-2pt$^{\prime\prime}$}

\newcommand{\lmd}{L$_\mathrm{bol}$-M$_\mathrm{env}$}
\newcommand{\mol}{Mol160/IRAS23385+6053}
\begin{document}
   \title{The pre-ZAMS nature of Mol160/IRAS23385+6053 confirmed by\textbf{ \textit{Spitzer}}}

   \subtitle{}

   \author{S. Molinari
          \inst{1}
          F. Faustini
          \inst{1}
          L. Testi
          \inst{2,3}
          S. Pezzuto
          \inst{1}
          R. Cesaroni
          \inst{2}
          \and
          J. Brand
          \inst{4}
%          \fnmsep\thanks{}
          }

   \offprints{}

   \institute{Istituto Fisica Spazio Interplanetario - INAF, Via Fosso del Cavaliere 100, I-00133 Roma, Italy \\
              \email{molinari, faustini, pezzuto@ifsi-roma.inaf.it}
		\and
             Osservatorio Astrofisico di Arcetri - INAF, Largo E. Fermi 5, I-50125 Firenze, Italy\\
             \email{lt, cesa@arcetri.astro.it}
		\and
			ESO, Karl-Schwarzschild-Strasse 2, D-85748, Germany \\ 
			\email{ltesti@eso.org}
        \and
			Istituto di Radioastronomia - INAF, Via Gobetti 101, I-40129 Bologna, Italy\\
             \email{brand@ira.inaf.it}
             }

   \date{Received ; accepted}

% \abstract{}{}{}{}{} 
% 5 {} token are mandatory
 
  \abstract  % context heading (optional)
{The formation of massive stars goes through phases which remain
heavily obscured until the object is well on the main sequence. The
identification of massive YSOs in different evolutionary phases is
therefore particularly difficult, and requires a statistical approach
with large samples of candidate objects to determine the observational
signatures of these different phases.}
% aims heading (mandatory)  d
{A mandatory activity in this context is the identification and
characterisation of all the phases that a massive forming YSO
undergoes.  It is of particular interest to verify the observability
of the phase in which the object is rapidly accreting while not yet
igniting the fusion of hydrogen that marks the arrival on the ZAMS.}
{One of the candidate prototypical objects for this phase  is \mol,
which has  been the subject  of  detailed studies  which confirmed
and strengthened  the possibility  that this  massive  YSO may  be
in a pre-Hot Core stage. We further investigate this issue by means
of \textit{Spitzer} imaging and spectroscopy in the 5-70~\um\ range.}
%results  heading (mandatory)  
{The  dense core of \mol,  which up to
now had only been detected at submillimeter and millimeter
wavelenghts, with only upper limits below 20~\um,  has been revealed
for the first time at 24  and 70~\um\ by {\it Spitzer}.  These
observations confirm the earlier assumptions that this object is
dominant at  far-IR wavelengths. The complete 24~\um -3.4~mm
continuum  cannot  be fitted  with  a standard  model  of  a Zero-Age Main-Sequence (ZAMS)
star embedded in  an envelope.  A simple  greybody  fit  yields  a
mass  of 220~\msun. The  luminosity is slightly in excess  of
3000~\lsun, which is a  factor of 5  less than previous  estimates
when only  IRAS fluxes were available between 20 and 100~\um. The
source is  under-luminous by the same factor with  respect to  UCH{\sc ii}
regions  or  Hot-Cores of similar  circumstellar mass, and simple
models show that this is compatible with an earlier evolutionary
stage. Spectroscopy between 5-40~\um\ shows  that the physical
conditions  are typical  of a photo-dissociated or photo-ionised
region, where the required UV illumination may be provided by some of
the other sources revealed at $\lambda\leq$24\um\ in the same star
forming region, and which can be plausibly modeled as moderately
embedded intermediate-mass ZAMS stars.}
% conclusions heading  (optional),  leave  it  empty  if  necessary  
{Our  results strengthen the suggestion that the central  core in \mol\ is 
a massive YSO  actively accreting from its circumstellar envelope and which 
did not yet begin hydrogen fusion.}

   \keywords{Stars: formation --  Stars: pre-main sequence -- Infrared: ISM}

   \authorrunning{Molinari et al.}
   \titlerunning{Pre-ZAMS nature of \mol}
   \maketitle
%
%________________________________________________________________

\section{Introduction}

The identification and the characterisation of all the different
phases that an intermediate and high mass forming object undergoes
during its approach toward the Main Sequence (MS) has received 
increasing attention in recent years. Besides its importance to
complete our understanding of the star formation process in general,
this research has interesting consequences in a broader context. For
example, the number of OB ZAMS stars indirectly estimated using a
variety of tracers like the radio continuum or the H$\alpha$ emission,
is widely used to estimate the star formation rate and -efficiency,
critical parameters to measure the history of star formation in 
our Galaxy as well as in external galaxies (McKee \&
Williams \cite{MKW97}, Kennicutt \cite{K98}).

%The definition of "protostar" for intermediate and high-mass objects
%is more elusive than for low-mass, since a considerable part of the
%luminosity comes from the contraction of the forming core even during
%a phase of intense accretion. We then favour a definition of a massive
%protostar as an object where the hydrogen burning has not yet been
%ignited, i.e. the YSO is not yet on the ZAMS. 
The identification of a "protostar" for intermediate and high-mass objects
is more difficult than for those of low-mass, because massive objects 
reach the ZAMS while still accreting material. Up to that moment they 
remain deeply embedded in their natal molecular cloud, and unlike their 
lower-mass counterparts they lack a visible pre-MS phase. 
With accretion rates
lower than 10$^{-5}$~\msunyr\ the existence of this classical pre-MS
phase is limited to objects of mass M$\leq$8~\msun. Accretion, however,
plays a very important role raising this mass limit and allowing the
arrival on the ZAMS at higher stellar masses if accretion proceeds at
much higher rates; the Young Stellar Object (YSO) joins the ZAMS at M=16~\msun\ for \mdot
=10$^{-4}$~\msunyr (Palla \& Stahler \cite{PS92}), and well beyond
20~\msun\ for even higher rates. Such rates seem indeed plausible,
given the mass loss rates measured from outflows emanating from
massive YSOs (Zhang et al. \cite{Zhang01}, Beuther et
al. \cite{Beu02}), and are also predicted by theoretical models (McKee
\& Tan \cite{MKT03}) where accretion rates increase with the mass of
the forming core.

We recently completed a preliminary study of the Spectral Energy 
Distributions (SEDs) of a sample of
massive YSOs in various evolutionary stages (Molinari et
al. \cite{Moli08}) which suggests that objects more massive than
8~\msun\ may indeed be observable in a pre-ZAMS phase as pure
far-infrared (undetected below 20~\um) objects underluminous with
respect to Hot-Cores or Ultra Compact (UC) H{\sc ii} regions for a given envelope mass.  
\mol\ clearly manifests its difference in multiwavelength
appearance with respect to known Hot-Cores or UCH{\sc ii} regions, 
and is perhaps a template for this class of massive analogues of 
classical Class~0 objects (Molinari et al. \cite{Moli98b}). 
The strong millimetric peak, which appears to harbour the driving source
of a molecular outflow, is anti-correlated with the complex emission
morphology visible at mid-IR wavelengths (Molinari et
al. \cite{Moli98b}) and with the extended radio continuum emission
found in the region (Molinari et al. \cite{Moli02}). Detailed
spectroscopic investigation confirmed that from a chemistry viewpoint,
the millimeter core of \mol\ appears to be on the verge of turning
into a Hot-Core (Thompson \& Macdonald \cite{TM03}) and, with
T$\leq$50K (Fontani et al. \cite{Fontani04}), is considerably colder
than typical Hot-Cores (HC) of similar envelope mass.

The suggestion that bolometric luminosity and SED morphology may be
promising evolutionary indicators is based on the coupled analysis of
mid-IR and submillimetric data. For the study of large samples in
these wavelength ranges, where survey data are available, the
accessible spatial resolutions of 10-20\asec\ are not sufficient to
pinpoint each and every emitter in the clustered environments where
massive YSOs are found, but are enough to recognize systematic
differences between classes of objects. Similar resolutions in the
far-infrared, where they are most crucial since it is where the SEDs
peak, will not be accessible until the Herschel satellite is 
available; in the meantime, an important confirmation of the scenario
proposed by Molinari et al. (\cite{Moli08}) may be provided by the
{\it Spitzer} satellite. Stecklum et al. (\cite{Steck05}) presented MIPS
continuum data of \mol\ and concluded that this source was not a
high-mass protostar. We here reconsider the same MIPS data and provide
an alternative and more plausible intepretation which is more 
consistent with
the independent evidence of the pre-HC nature of this object and the
more general picture of the luminosity evolution of massive
YSOs. Additional {\it Spitzer} IRS spectroscopy is also presented, and the
implications for the star formation activity of the entire region are
discussed.

\section{Observations and Data Analysis}
\label{obs}

\begin{figure*}[ht]
\centering
\resizebox{\hsize}{!}
{\includegraphics[angle=-90,origin=c,width=8cm]{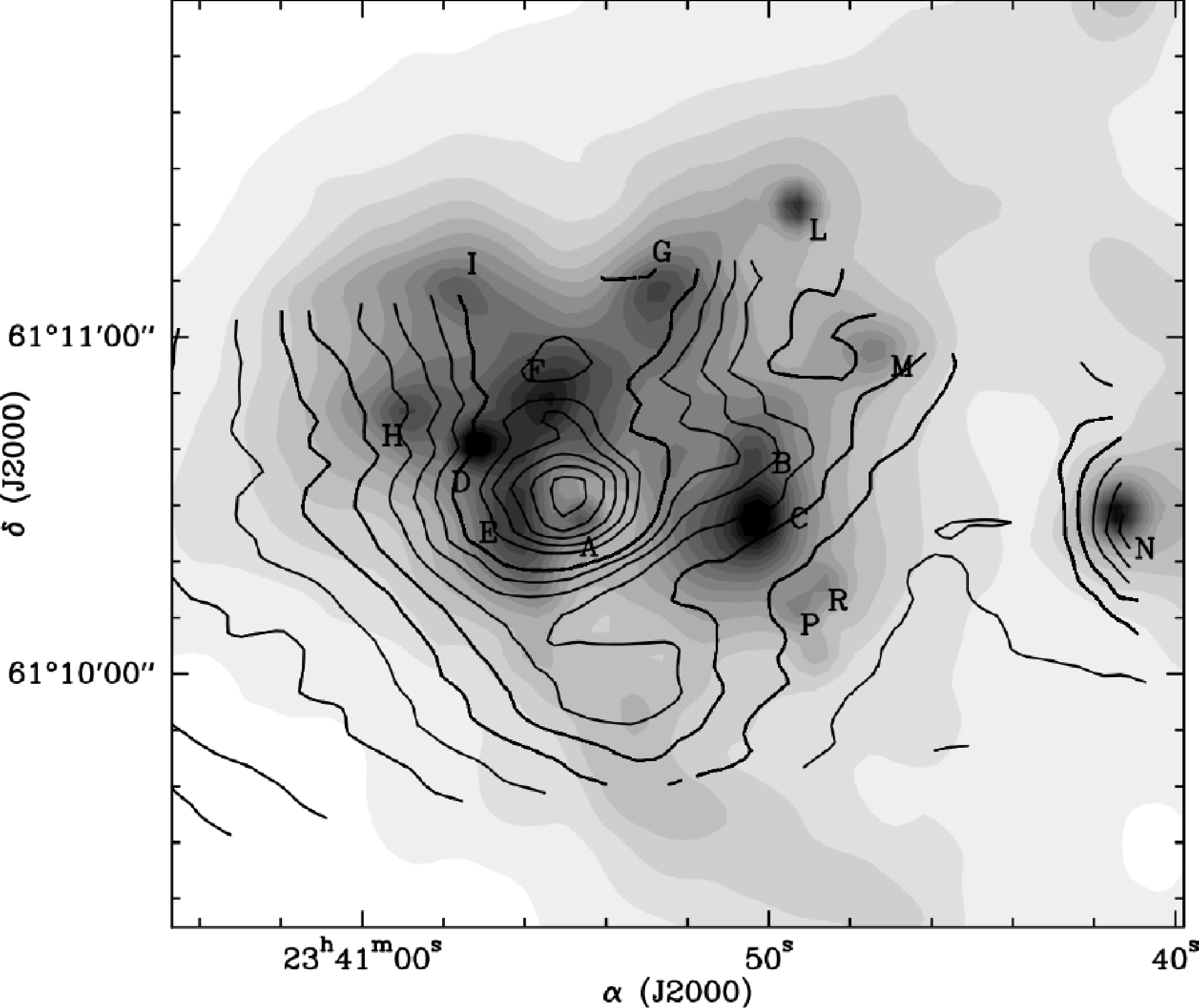} 
\includegraphics[angle=-90,origin=c,width=8cm]{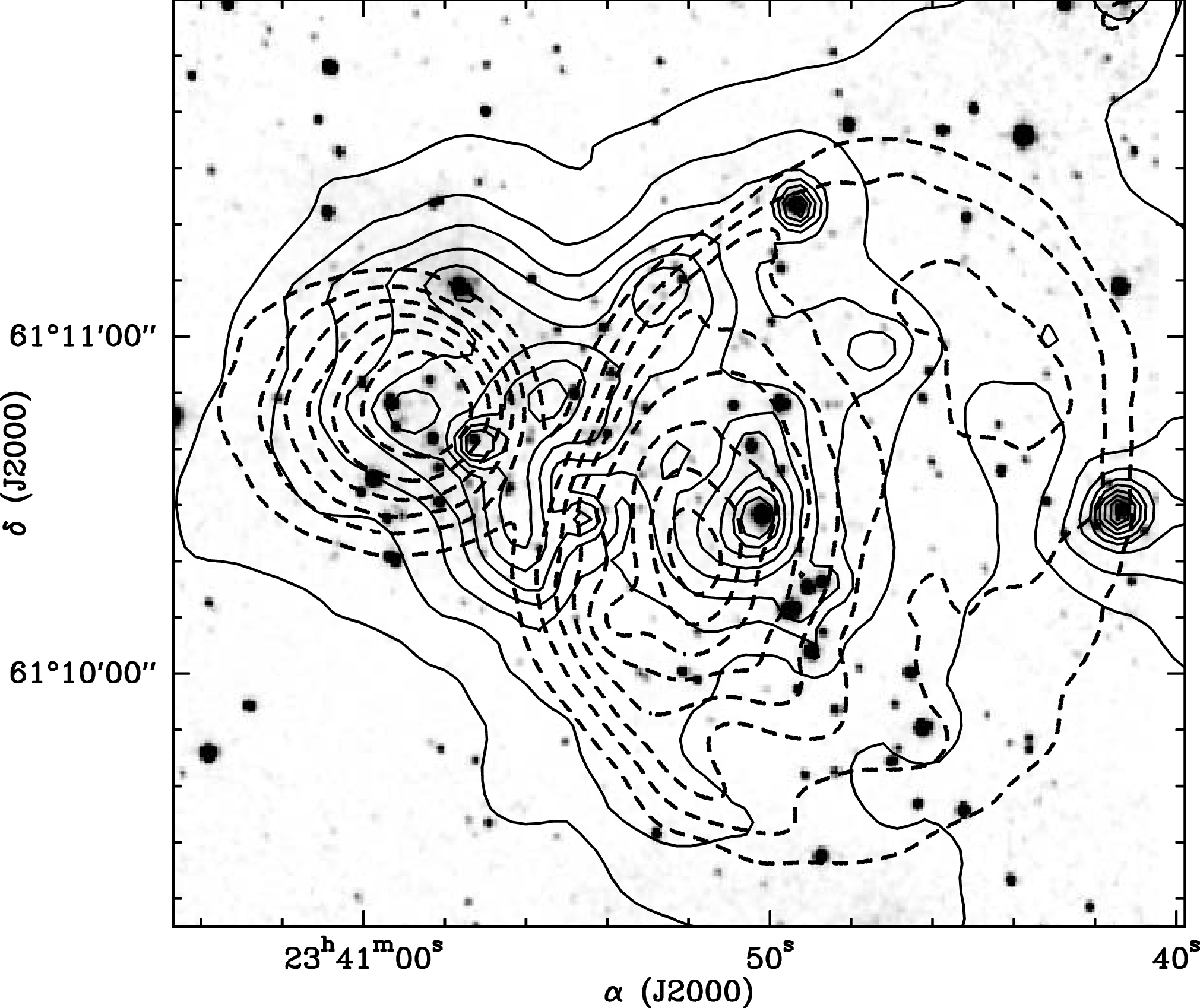}}
\caption{\textbf{a)} (Left) MIPS 24~\um\ grey-scale image of the \mol\
field, with superimposed the contours of the 70~\um\ emission; the
letters mark the identified sources. \textbf{b)} (Right) 2.2~\um\
image of the same region, at approximately the same scale; solid
countours represent the 24~\um\ continuum emission, while the dashed
contours show the 3.6~cm thermal free-free emission.}
\label{mipsfig}
\end{figure*}

Data for source \mol\ were acquired with the {\it Spitzer} satellite in two
Guaranteed Time programs. The first program used the MIPS instrument
(Rieke et al. \cite{Rieke}) to acquire photometric images at 24 and
70~\um. The $128\times 128$ Si:As array at 24~\um\ has a pixel size of
2\pas49 $\times$ 2\pas60 and a field-of-view (FOV) of 5\pam4 $\times$ 
5\pam4, while the 70~\um\ array consists of $32\times 32$ Ge:Ga matrix 
with a 
pixel size of 5\pas24 $\times$ 5\pas33 and a FOV of 2\pam7
$\times$ 1\pam35. The data were taken with an exposure time of 3~sec, and
were repeated twice at 24~\um\ and 3 times at 70~\um. One cycle of the
basic 24~\um\ photometry observation consisted of 14 offset images of
the field combined with classical techniques of chopping (along-scan
mirror motion) and dithering. One cycle of 70~\um\ observations
produces, with the same method, 12 images of the field. Also an
estimate of background-foreground contribution is made acquiring an
image off-source with chopping technique.

The second program contains spectroscopy mapping with the IRS
instrument (Houck et al. \cite{Houck}) using three of the available
modules. The "Long High" (LH) module covers a range between 19 and
37~\um\ at a spectral resolution of R=600 with a FOV of 22\pas3
$\times$ 11\pas1 and a pixel size of 4\pas5. The "Short High" (SH) module
covers with similar spectral resolution the range between 10 and
19.5~\um\ with a FOV of 11\pas3 $\times$ 4\pas7 and a pixel size of
2\pas3. The "Short Low" (SL) slits were used to cover the range
between 5.6 and 14.5~\um. The SL module is composed of two sub-modules
at different wavelengths; one covers the range between 7.4 and
14.5~\um\ at R=16, with a FOV of 57\asec$\times$3\pas7 and a pixel
size of 1\pas8, and the second covers at R=8 a range between 5.2 and
8.7~\um, with a FOV of 57\asec$\times$3\pas6 and a pixel size of
1\pas8. Observations consisted of one integration with a ramp
duration of 6 seconds. The area around the position of the millimeter
emission peak was covered with a small $3\times 2$ map.

For both programs we downloaded the Basic Calibrated Data products from the 
{\it Spitzer} archive. Pipeline versions used for automatic processing 
were S.11.0.2 for MIPS data, and S15.3.0 for IRS.

\subsection{Source photometry: Near-IR to millimeter}
\label{mipsobs}

Fig.~\ref{mipsfig}a shows the \mol\ field at 24~\um\ (greyscale image)
and at 70~\um\ (contours). The extraction and photometry of point
sources from the 24~\um\ image was attempted using several publicly
available packages for automatic source detection and photometry
(SExtractor, IRAF, MOPEX). As is apparent from the image, the
various peaks that are visible sit on a complex patch of extended emission
which is strongly variable on small and large scales. This poses
severe problems for reliable source detection and photometry since
the various packages we tried assume either a constant background, or
estimate one by using a box of fixed size across the
image. Sources like A in the figure, for example, are undetected with
SExtractor and also with MOPEX if the box for estimating the background is 
too large. Using a small box would solve this, but on the other hand leads
one to significantly understimate the fluxes for sources like N or E
which instead lie on a less variable and broader patch of extended
emission; discrepancies in the fluxes of up to a factor of 5 are found for 
some objects using different packages. For this reason we decided to
visually identify compact peaks in the map, and to manually perform 
the photometry at 24~\um\ with custom procedures using the GILDAS
package, optimising by hand the area for the background estimate.

\begin{table*}[ht]
\setlength{\tabcolsep}{0.1in}
%\begin{flushleft}
\caption{Photometry of sources in the field}
\begin{tabular}{lccccccccccccc}\hline\hline
Source & & $\alpha$(J2000) & $\delta$(J2000) & F$_B$ & F$_R$ & F$_J$ & F$_H$ & F$_{K_s}$ & F$_{6.75\mu m} ^{a}$ & F$_{15\mu m} ^{a}$ & F$_{24\mu m}$ & F$_{70\mu m}$ & F$_{850\mu m}$ \\
 & & & & mJy & mJy & mJy & mJy & mJy & Jy & Jy & Jy & Jy & Jy\\ \hline
A &  & 23:40:54.65 & +61:10:27.6 & $-$ & $-$ & $-$ & $-$ & $-$ & $-$ & $-$ & 0.14 & 55 & 2.2 \\
B &  & 23:40:50.42 & +61:10:37.4 & $-$ & $-$ & 0.8 & 1.5 & 1.7 & 0.2  & 0.14 & 0.16 & $<$25 & $<$0.3 \\
C &  & 23:40:50.25 & +61:10:27.6 & $-$ & $-$ & 1.5  & 6.1 & 10.6 & $<$0.02 & 0.6 & 0.6 &  $<$25 & $<$0.3 \\
D &  & 23:40:57.11 & +61:10:41.0 & $-$ & $-$ & 0.07 & 0.35 & 0.7  &  0.2  & 0.1 & 0.3  & $<$25 & $<$0.4 \\
E &  & 23:40:56.35 & +61:10:30.0 & $-$ & $-$ & 0.07  & 0.16  & 0.6 & 0.3  &  0.2 & 1.0 & $<$35 & $<$0.8 \\
F &  & 23:40:55.50 & +61:10:49.0 & $-$ & $-$ & $-$ &  $-$ &  $-$ &  0.26 &  0.2  & 0.5 & 12 & $<$0.5 \\
G &  & 23:40:52.62 & +61:11:09.2 & $-$ & $-$ & 0.1 & 0.06 & 0.2 &  $-^e$  & $-^e$  & 0.13 & $<$30 & $<$0.5 \\
   & {\,\,\,\,\,1} & 23:40:58.97 & +61:10:27.9 & 0.1 & 0.4 & 0.07 & 0.5 & 0.5 &  & & & & \\
H$^b$ &  \multirow{-2}{0.25cm}{\Huge{\{}}2&  23:40:59.04 & +61:10:33.3 & 0.05 & 0.2 & 0.07 & 0.1 & 0.1  & $<$0.01 & $<$0.15 & 0.14 & $<$20 & $<$0.2 \\
      & {\,\,\,\,\,3} & 23:40:59.11 & +61:10:43.8 & $-$ & $-$ & 0.8 & 0.8 & 0.6 & & & & & \\
I     & &  23:40:57.62 & +61:11:08.0 & 0.2 & 0.7 & 6.4 &  6.6 & 5.1 & 0.24 & 0.1 & 0.08  & $-^c$ & $<$0.2 \\ 
L    & &  23:40:49.23 & +61:11:23.9 &  $-$ & $-$ & 0.64 & 2.0 & 5.5 & $-^e$ & $-^e$ &  0.3 &  $-^c$ & $<$0.1 \\ 
M   &  &  23:40:47.45 & +61:10:57.6 & $-$ & $-$ & $-$ & $-$ & $-$ & 0.04 & 0.01 & 0.33 & $<$15 & $<$0.2 \\
N    & &  23:40:41.44 & +61:10:28.1 & $-$ & $-$ & 0.5 & 1.5 & 2.4 &  $-^e$ & $-^e$  &  0.4  & $-^c$ & $-^d$ \\
P    & &  23:40:49.23 & +61:10:11.6 & $-$ & $-$ & 0.4 & 1.7  & 4.9  & 0.02 & 0.03 & 0.1 & $<$15 & $<$0.4 \\
R    & &  23:40:48.55 & +61:10:15.9 & $-$ & $-$ & 1.0 & 1.3 & 2.1 &  $-^e$  & $-^e$ & 0.08  & $<$15 & $<$0.4 \\
\hline
\end{tabular} 
%\end{flushleft}
\parbox{18cm}{$^a$ from ISOCAM images (Molinari et al. \cite{Moli98b})}\\
\parbox{18cm}{$^b$ Although the positional association of other K$_s$ sources 
to the 24~\um\ peaks is very good (see Fig. \ref{mipsfig}b), it seems likely 
that source H, since it is much shallower than the other 24~\um\ peaks, could 
be associated to one of the three nearest K$_s$ sources visible in its close 
proximity.}\\
\parbox{18cm}{$^c$ This source is not covered by the 70~\um\ image.}\\
\parbox{18cm}{$^d$ This source is not covered by the 850~\um\ image.}\\
\parbox{18cm}{$^e$ This source is not covered by the ISOCAM images.}
\label{fluxtab}
\end{table*}

\begin{figure*}[ht]
\centering
\resizebox{\hsize}{11cm}{\includegraphics{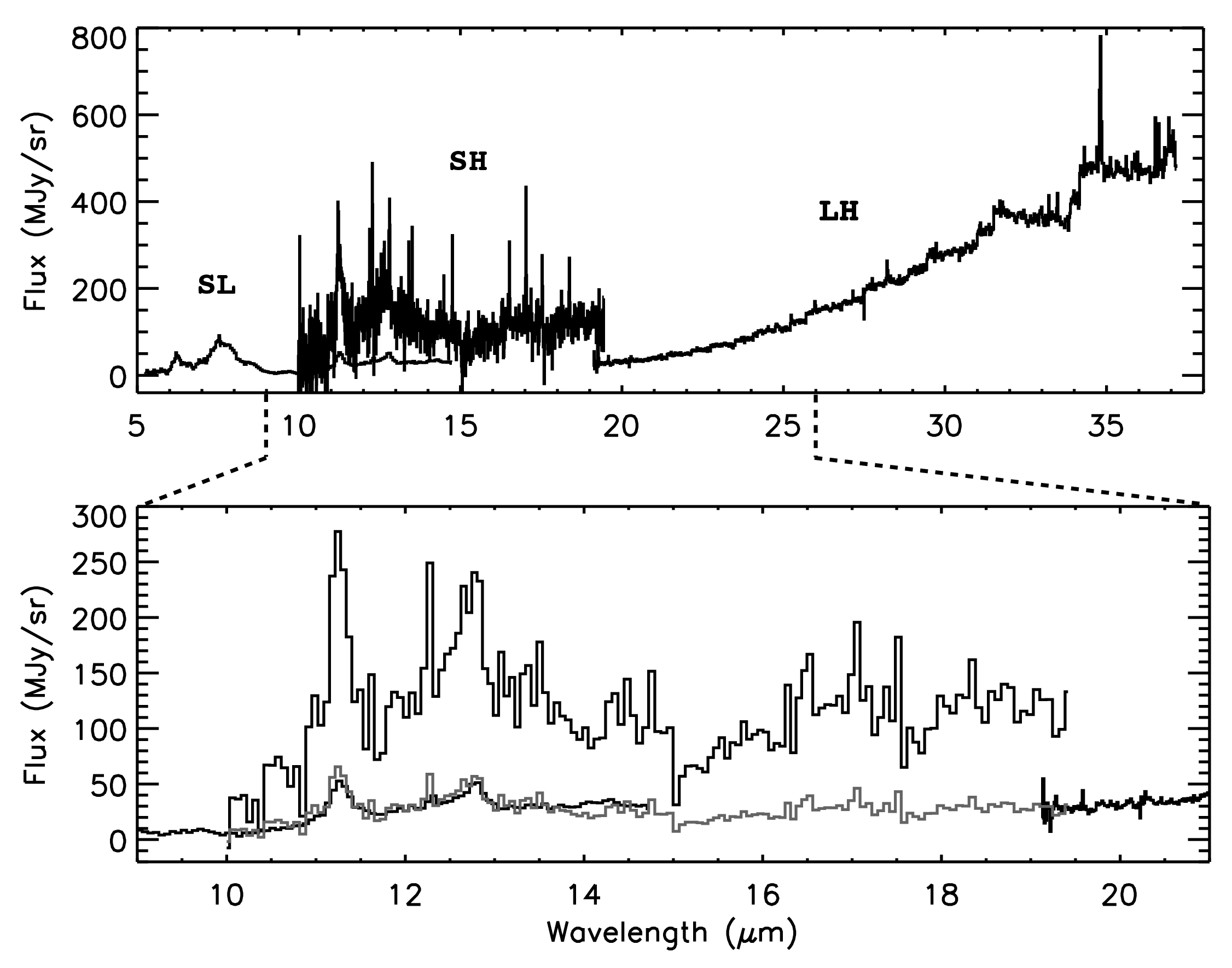}}
\caption{Full IRS spectrum (top panel) extracted using CUBISM at the position 
of the central millimeter core (source~A). The bottom panel is an enlargement 
where this time the SH module spectrum is resampled at the resolution of the 
SL module; its rescaling by a factor of 4.2 brings it to an almost perfect 
overlap with the SL spectrum, and joins very well with the LH spectrum at 
longer wavelength.}
\label{m160irsfig}
\end{figure*}

At 70~\um\ the situation is less complex because due to the lower
spatial resolution the various sources in the area are blended in a
large patch of emission; among the sources detected at 24~\um\ source~A is 
the one that more clearly stands out above this plateau 
(Table~\ref{fluxtab}). A slight offset is apparent between the position of 
the 24 and 70\um\ peaks; we will comment on this below. The size of this 
peak was estimated analyzing
cross-cut brightness profiles; this information was transferred to
the AIPS task JMFIT to fit a composite 2D-Gaussian and a plateau
resulting in an integrated flux density of 65~Jy for the compact
central peak coincident with source A. The integrated intensity of the
entire 70~\um\ emission structure, after subtraction of a constant
emission level clearly detectable above the noise at the borders of
the image, is $\sim 350\div 400$~Jy, compatible with the 60~\um\ IRAS
flux density. The 70~\um\ flux density that can be assigned to the
central massive core coincident with source A is then much lower than
the 60~\um\ IRAS flux density, contrary to the assumptions made by
Fontani et al. (\cite{Fontani04}). In that work we carried out a
far-IR extrapolation of the mid-IR emission of the bright patches
revealed in the ISOCAM images (Molinari et al. \cite{Moli98b}), which 
are also visible in the present 24~\um\ image, and concluded that most of
the IRAS 60 and 100~\um\ flux densities could be assigned to the
central core. The MIPS observations, however, provide a direct
measurement and show that only a minor fraction of the total
$\lambda\leq 70$~\um\ flux density in the region can be assigned to
the central core.

In addition to source~A, which is clearly dominant at 70~\um, we
detect a fainter peak which barely emerges from the plateau in
correspondence to source~F; as for source~A, the integrated flux is
estimated using the AIPS JMFIT task fitting a peak-plateau
combination. Another secondary peak is visible both at 24 and 70~\um\
about 35\asec\ South/South-East of A, but is considerably shallower
and broader than the other sources and is not considered in this
analysis. As for the other sources detected at 24~\um, we
assigned the local values of the 70~\um\ emission as upper limits.

The 24-70~\um\ spectral region alone, although very important as the
SED rises toward the peak for typical YSOs envelopes, is not
sufficient for a reliable comparison with SED models. Leaving aside
IRAS whose beam at 100~\um\ encompasses more than the entire area
shown in Fig.~\ref{mipsfig}a, the longer wavelength information is
extracted from submillimeter and millimeter images (Molinari et
al.~\cite{Moli02}). Similarly to the MIPS 70~\um\ image, the 850~\um\
SCUBA image shows an extended emission patch covering the entire area,
in which a strong peak coincident with source~A clearly stands out;
850~\um\ fluxes for the various 24~\um\ sources were assigned
following the same approach as for the 70~\um\ fluxes. The peak at
source~A is also exactly coincident with the core visible in the 
interferometric OVRO
images at 3.4~mm; the slight offset of the 70\um\ peak position with respect 
to the 24\um\ source A-peak may then not be real and could be ascribed to an 
inaccuracy of the 70\um\ image astrometry. 

Concerning the shorter wavelengths, for most of the sources we could
estimate flux densities at 6.75 and 15~\um\ from ISOCAM images
(Molinari et al. \cite{Moli98b}) using the same approach followed for
the MIPS 24~\um\ image; sources G, L, N  and R are outside the ISOCAM
FOV. Near-IR counterparts in J, H, and K$_s$ could also be assigned for
most of the sources (Faustini et al., A\&A submitted). The K$_s$-image is
shown in greyscale in the right panel of Fig. \ref{mipsfig}; the
full-line contours report the 24~\um\ emission, while the dashed
contours mark the emission pattern in the 3.6~cm thermal free-free
emission (Molinari et al.~\cite{Moli02}). The only dubious counterpart
assignment concerns source H, for which there are three NIR objects
very close to the 24~\um\ peak; we list all of them in
Table~\ref{fluxtab}.

For those 24~\um\ sources with NIR counterpart we tried to extend the
SED coverage to the visible range by visual inspection of the Digital
Sky Survey 2 blue and red plates. Where counterparts could be found,
fluxes were estimated doing standard aperture photometry and
calibrating the integrated flux using a conversion factor estimated
from the magnitude of medium intensity sources in the field as
reported in the USNO-B1 catalogue.  Zero-magnitude fluxes of 4260~Jy
and 3080~Jy for the B and R bands were used (Bessel
\cite{Bessel79}). 

\subsection{IRS Spectroscopy}
\label{irsobs}

\begin{figure*}[ht]
\centering
\resizebox{\hsize}{11cm}{\includegraphics{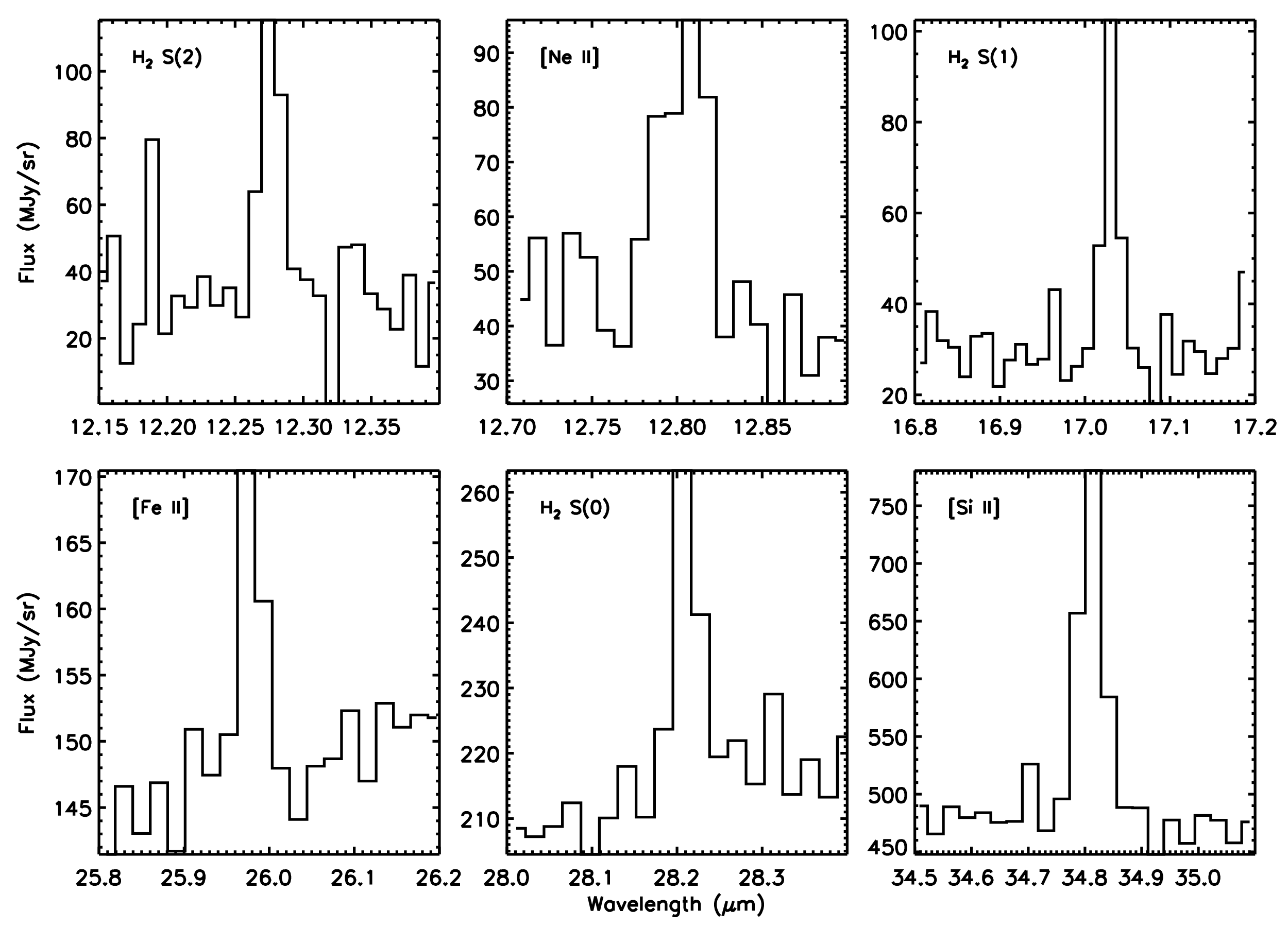}}
\caption{The six identified lines in the entire spectrum at the position of 
source~A, as extracted from the spectrum in Fig.~\ref{m160irsfig}; the various 
lines are identified in each panel.}
\label{m160irslines}
\end{figure*}

The areas observed in the $3\times 2$ spectroscopic map are different across
the wavelength range since the slits in the different modules have
different sizes and orientations; however they are much smaller than
the area imaged by MIPS. All slits were centered on the millimeter
peak position (coincident with source A), so that we have complete
spectral coverage for this source. We used the CUBISM software (Smith
et al. \cite{Smith07}) to mosaic and co-add the spectra from all the
observed positions and obtain convenient spectral cubes for further
analysis. We present in Fig. \ref{m160irsfig} the complete spectrum in
surface brightness units extracted at the position of source A (top
panel).

Spectra from the two SL1 and SL2 modules for $\lambda \leq$15\um\ have
been merged together and clearly show a set of Poly-Aromatic
Hydrocarbon (PAH) features at wavelengths between 6 and
13~\um. The SH spectrum is particularly noisy not only with respect to
the lower resolution SL portion, but also with respect to the
LH. Besides, it is also apparent (see bottom panel of
Fig. \ref{m160irsfig}) that the SL and LH portions appear to be quite
well aligned among them, while the SH portion seems anomalous in this
respect. To understand the nature of the displacement of SH we
resampled it to the same resolution as the SL portion, with which SH
overlaps between 10 and 15\um; the resampled SH is shown in black in
the bottom panel of Fig. \ref{m160irsfig}; normalising it to the SL
over the entire overlapping range we obtain a ratio of 4.2, and if we
rescale SH by this factor (the grey line in the figure) we see that
SH almost perfectly overlaps with SL. We conclude that the SH
spectrum likely suffers from an incorrect calibration, and a "gain"
type of correction brings it in excellent agreement with both SL and
LH. We then rescale SH by this gain factor, and we will use this
rescaled spectrum to extract lines and estimate integrated fluxes.

Figure \ref{m160irslines} shows six lines identified in the SH and LH
spectra. \neii\ line and the S(1) and S(2) pure rotational lines of
H$_2$ are extracted from the SH spectrum; the LH spectrum is
relatively less noisy and allows us to reliably identify the \feii,
the H$_2$ S(0) and the \sii\ lines. Integrated fluxes were derived 
using Gaussian fits to line profiles after subtraction of linear baselines; 
estimated values are reported in Table~2 for the position of source~A.

\begin{table}
\begin{flushleft}
\caption{Integrated line fluxes at the position of source~A.}
\begin{tabular}{lcc} \hline\hline
Line & Wavelength & Flux \\
        & (\um)     & W cm$^{-2}$ sr$^{-1}$ \\ \hline
H$_2$ (0-0)S(0) & 28.2 & $6.4\times 10^{-13}$ \\
H$_2$ (0-0)S(1) & 17.02 & $1.7\times 10^{-12}$ \\
H$_2$ (0-0)S(2) & 12.28 & $3.2\times 10^{-12}$ \\
{[Ne\sc{ii}]} & 12.8 &  $3.2\times 10^{-12}$ \\
{[Fe\sc{ii}]} & 26 &  $3.5\times 10^{-13}$ \\
{[Si\sc{ii}]} & 34.8 & $4.3\times 10^{-12}$ \\ 
\hline
\end{tabular}
\end{flushleft}
\label{irslines}
\end{table}

\section{Source~A: an accreting pre-ZAMS massive YSO}

Source~A is the most noticeable object in this region. While
undetected in the near-IR and ISOCAM images, it is first visible at
24~\um\ and then becomes the most prominent source in the far-IR, the
submillimeter and the millimeter regions, and it coincides with the
location of the driving source of the molecular outflow seen in \hcop\
and \sio\ (Molinari et al.~\cite{Moli98b}).

\begin{figure}
\centering
\resizebox{\hsize}{!}{\includegraphics[angle=90]{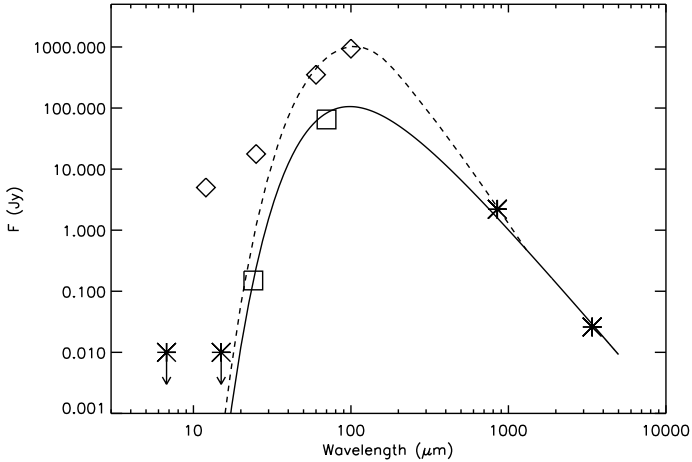}}
\caption{Spectral Energy Distribution (SED) for the central core of \mol. The 
diamonds are the IRAS fluxes, the squares are the MIPS fluxes and the asterisks 
represent the ensemble of measurements from ISOCAM (upper limits since there 
is no detection below 24~\um), SCUBA and OVRO; the dashed line is a greybody fit to the above data set using the IRAS 60 and 100~\um fluxes (as in Molinari et 
al.~\cite{Moli98b}). The full line is a greybody fit 
to the SED using the MIPS data instead of the IRAS ones.}
\label{m160sed}
\end{figure}

Source~A, however, is not as prominent in the far-IR
(Fig.~\ref{m160sed}) as we originally believed when only IRAS data
were available (Fontani et al.~\cite{Fontani04}). This has very
important consequences regarding its nature.

In a systematic analysis of the SEDs of a sample of 42 sites of massive star
formation (Molinari et al.~\cite{Moli08}) we used the radiative
transfer models of Whitney et al. (\cite{Whitney03}) to compute a grid
of SEDs for embedded ZAMS stars of various spectral classes (from B5
to O3) and for a wide range of envelope properties (radius, mass,
geometry). In that work the far-IR portion of the SED was based on
IRAS fluxes (the diamonds in Fig. \ref{m160sed}), 
corrected for a possible contribution from extended emission.
We could not provide a successful fit with any of
the models of our grid for what we now call source A in \mol.
Likewise, using the much more reliable
estimate of the source SED in the far-IR range that the MIPS data
allow (the squares in Fig. \ref{m160sed}), we still cannot find any
acceptable fit with models of embedded ZAMS stars; any model that fits
the 850~\um\ and 3.4~mm points predicts far too much flux
shortward of 100~\um. We neglected the possible contribution of 
free-free emission to the 3.4mm flux as this could amount to a 10-20\% level 
at most, given that source A was not detected at 2cm at the VLA in the 
B-configuration (Molinari et al. \cite{Moli98a}), and in the extreme case of 
optically thick free-free (F$\propto \nu ^2)$ up to the millimeter range.

The full line in Fig.~\ref{m160sed} represents
the best fit, using a simple greybody which assumes constant density
and temperature. The fit uses the OVRO, SCUBA, and MIPS data, and the 
ISOCAM upper limits
(the asterisks) but neglects the 12 and 25~\um\ IRAS data since they
are clearly due to the other YSOs in the region (see
Fig.~\ref{mipsfig}). The overall fit is remarkably good and
corresponds to an envelope with T=37~K, M=220~\msun\ and a dust
opacity index $\beta$=1, assuming a dust opacity of
0.005~cm$^2$g$^{-1}$ at 1.2~mm and a gas-to-dust ratio of 100
(Preibisch et al.~\cite{Prei93}).

Robitaille et al. (\cite{Robi06}) used the same radiative transfer code 
of Whitney et al. (\cite{Whitney03}) that we use in Molinari et al. 
(\cite{Moli08}), to produce a much more extensive grid of SED models 
(not yet available at the time of our previous work) including central 
stars in all pre-MS phases from the birthline to the ZAMS. It is
quite convenient to use the automatic SED fitting tool provided by
Robitaille et al. (\cite{Robi07}), which reports the best fitting
models in increasing $\chi^2$ order. Fitting the SED of source A with 
this grid of models provides a best fit for a
central object of $\sim$9.5~\msun\ with a radius of $\sim$25~\rsun\
and a surface temperature slightly in excess of 8000~K accreting at 
$\sim 2\cdot$10$^{-3}$~\msunyr\ from a massive $\sim 500$~\msun\
envelope. This value of the stellar radius is much higher than typical
values for ZAMS stars of that mass; besides, with these accretion
rates the stellar mass needed to ignite deuterium burning is higher
than 20~\msun\ (Palla \& Stahler~\cite{PS92}), so that such an object
would still be in the pre-ZAMS phase. Clearly, given the number of
parameters involved in the modelling and the limited number of data
points available, the fitting cannot pinpoint a unique combination of
parameters but rather identifies a certain range in the parameter 
space. It is true, however, that the possibility of an embedded ZAMS
star seems to be reliably excluded and that the above-mentioned
combination of parameters is representative of the region of the
parameter space identified by the SED fitting.

The greybody fit represented by the full line in Fig.~\ref{m160sed}
provides an integrated bolometric luminosity of 3170~\lsun\ at a
distance of 4.9~kpc (similar to what was obtained by Stecklum et
al. \cite{Steck05}). This is remarkably lower than the
$\sim$16\,000~\lsun\ obtained again fitting a greybody but using the
IRAS PSC 60 and 100~\um\ fluxes (the dashed line in
Fig.~\ref{m160sed}, see also Molinari et al.~\cite{Moli98b}), or than
the $\sim$10\,000~\lsun\ obtained if we try to correct the far-IR
fluxes for extended emission using higher spatial resolution
information from the submillimeter, as done in Molinari et
al. (\cite{Moli08}). Thanks to the MIPS data we can now set much more
stringent constraints on the SED over the entire wavelength range and
we can obtain this much lower value for the luminosity while the 
estimated envelope mass remains basically the same.

\begin{figure}[ht]
\centering
\resizebox{\hsize}{!}{\includegraphics[angle=90]{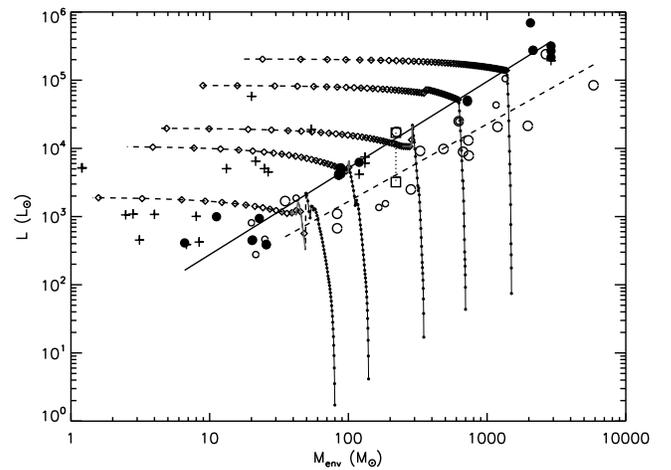}}
\caption{\lmd\ diagram from Molinari et al. (\cite{Moli08}); full 
circles are IR-P sources, empty circles are MM-P sources and 
plusses are for IR-S sources. Dotted full lines and diamond dashed lines 
are evolutionary models; the two squares connected by the vertical dotted 
line represent the extent of the shift of 
source~A due to the new luminosity estimate. The full and dashed diagonal lines represent a linear fit to the position of the IR-P and MM-P sources, respectively.}
\label{lmdiagm160}
\end{figure}

This decrease in the luminosity estimate has important consequences
concerning the evolutionary stage of the massive forming object. In
our recent analysis of the SED evolution of massive YSOs (Molinari et
al.~\cite{Moli08}) we could differentiate two classes of objects
depending on whether the entire SED from the mid-IR to the millimeter
can be consistently fitted with a single model of an embedded ZAMS
star (called ``IR-P'' by Molinari et al. and represented with the
filled circles in Fig.~\ref{lmdiagm160}) or not. If not, the SED was
modeled with two components: a single-temperature greybody for the
millimetric/far-IR part (``MM-P'', open circles in
Fig. \ref{lmdiagm160}), and a mildly-obscured ZAMS star for the mid-IR
(``IR-S'', plus signs in Fig.~\ref{lmdiagm160}). These
different classes of objects occupy distinct regions in the \lmd\
diagram (Fig. \ref{lmdiagm160}), and a simple toy-model for the
evolution of the bolometric luminosity and the circumstellar envelope
can explain the sequence MM-P/IR-P/IR-S in evolutive terms analogous
to the Class~0-I-II sequence established for low-mass YSOs. From a
statistical viewpoint the IR-P objects are dominated by sources with
far-IR colours of Hot Cores/UCH{\sc ii} regions, therefore 
confirming their ZAMS nature; this
is not the case for MM-P objects. In addition to the SED differences,
Fig.~\ref{lmdiagm160} shows that the luminosity is a critical
parameter that can trace with a high dynamical range the transition
between MM-P and IR-P objects.

The two squares in Fig.~\ref{lmdiagm160} illustrate the position of
the source with the luminosity that was found in our original work on
this source (Molinari et al.~\cite{Moli98b}) and the one determined
using the MIPS observations. Quite remarkably, the shift to lower
luminosity brings the source from the locus of the IR-P objects (the
full straight line, which in our interpretation marks the arrival onto
the ZAMS) to the locus of the MM-P objects (the dashed straight
line). In evolutionary terms within the framework of the model
proposed in Molinari et al. (\cite{Moli08}), this luminosity
correction makes the source more than 5\,10$^4$~years younger than we
have previously estimated; this is the time it takes, according
to the prescriptions of accelerating accretion models (McKee \&
Tan~\cite{MKT03}) to bring source~A onto the ZAMS from its present
location in the \lmd\ diagram. Source~A is even younger if we adopt
the $\sim 500$~\msun\ envelope mass derived from the detailed grid of SED
models of Robitaille et al. (\cite{Robi06}).

\begin{figure*}
\centering
\resizebox{\hsize}{!}{\includegraphics{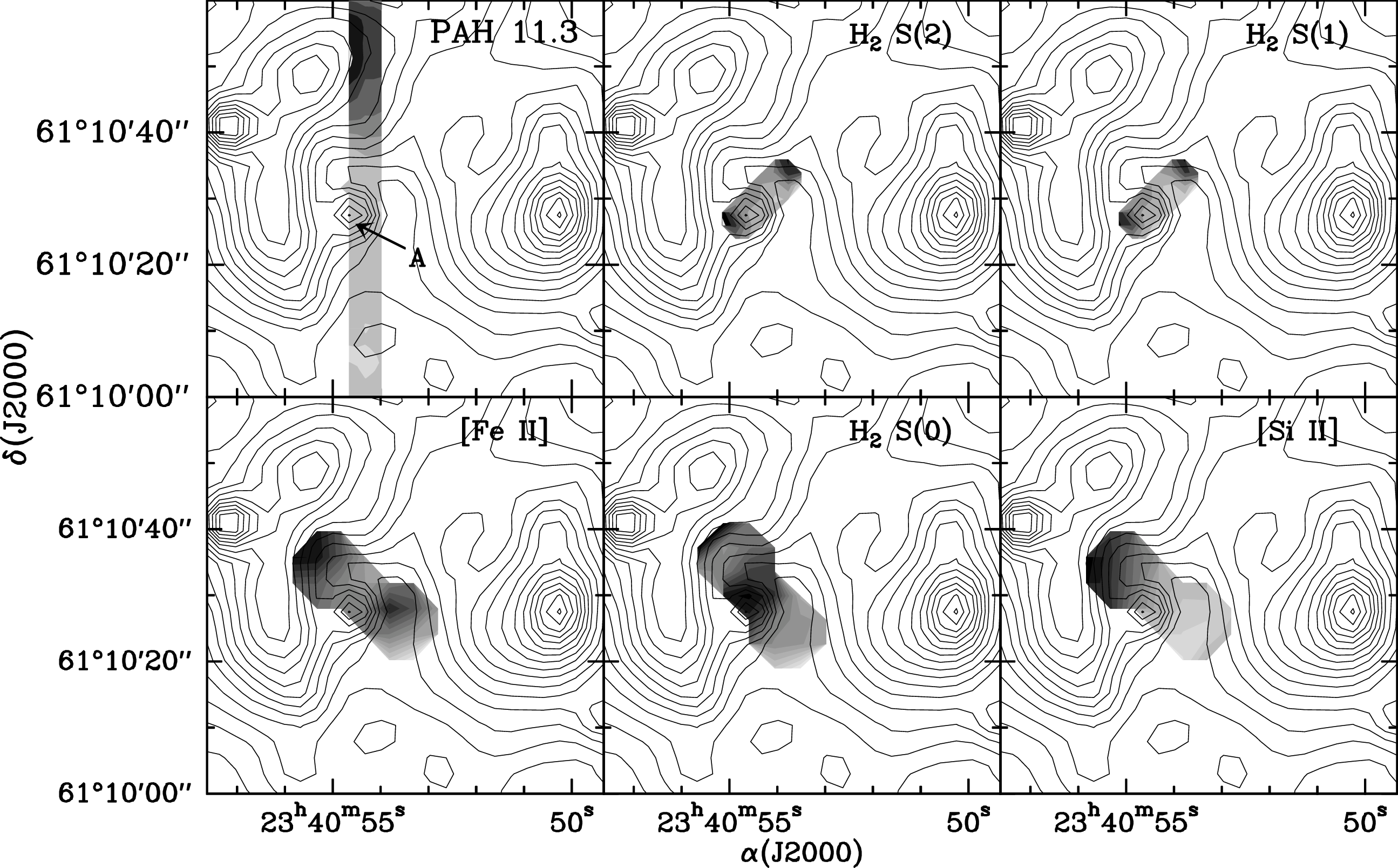}}
\caption{Grey-scale maps (intensity increases from white to black) of  
continuum-subtracted integrated line intensities superimposed on the the 
contours of the 24~\um\ continuum emission. The various panels show, from 
left to right and from top to bottom: the PAH 11.3~\um\ feature, H$_2$ S(2), 
H$_2$ S(1), \feii, H$_2$ S(0), and \sii. The map for the \neii\ line is not 
reported but the emission distribution is virtually identical to the 
H$_2$ S(1) and S(2) lines.}
\label{m160irsmaps}
\end{figure*}

Molinari et al. (\cite{Moli08}) also suggested that MM-P objects are 
characterized by a much
steeper SED for $\lambda\leq 100$~\um\ compared to IR-P objects. In
particular we predicted that the [24-70] colour (estimated by 
interpolating observed mid-IR and sub-mm spectra using SED modeling) 
to be $\sim$1 for IR-P objects and $\sim$4 for MM-P objects: the MIPS
observations provide a direct measurement of the SED shape in this
critical region: with [24-70]=2.6 for source~A, these measurements
strenghten the foundations for the working assumption in our previous
work.

\section{Other YSOs and the UV-field in the region}
\label{uv}

Since source~A is only detected longward of 24~\um\ in the continuum,
dust extinction will prevent any line emission at shorter wavelengths,
originating from the embedded YSO, to emerge and be
detectable. Therefore all PAH features, the \neii, and the two S(2)
and S(1) lines cannot originate from the massive central core; indeed
all emission features are detected with variable intensities
throughout the regions mapped with the IRS. To better investigate
their spatial distribution we again used CUBISM to produce
continuum-subtracted spatial maps of the integrated line fluxes. They
are presented in Fig.~\ref{m160irsmaps} for the 11.3~\um\ PAH feature
and for five of the detected lines, as indicated.

\begin{figure*}[t]
\centering
\resizebox{\hsize}{11cm}{\includegraphics{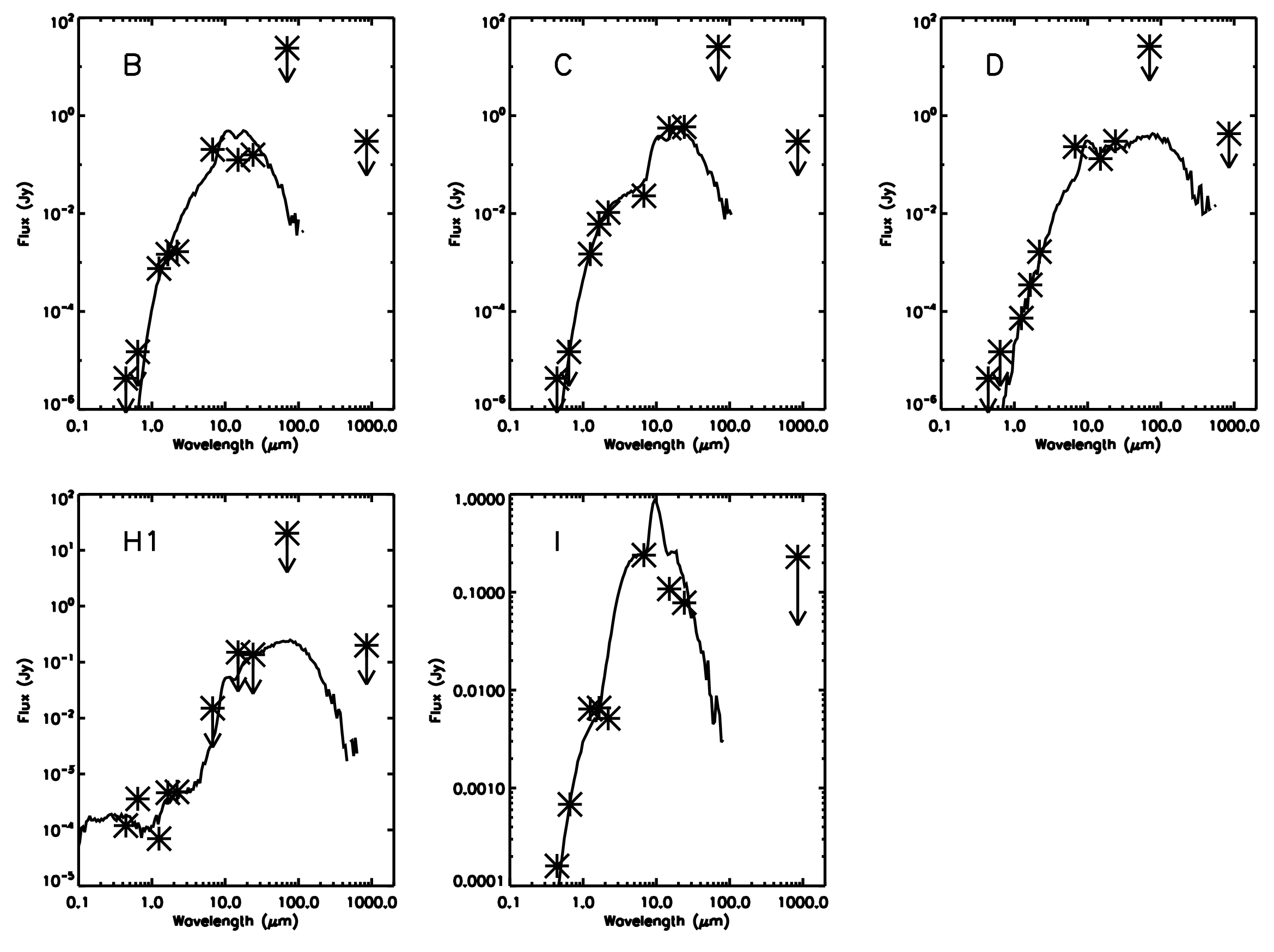}}
\caption{SEDs for the various sources where a fit with a model of an 
embedded ZAMS star could be found.}
\label{othermodels}
\end{figure*}

The intensity of the 11.3~\um\ feature (the same spatial distribution
is followed by all other PAHs) has a minimum at the location of source~A 
and increases towards the north where source F is located. PAHs emit in 
a non-equilibrium regime
in the presence of relatively energetic incident radiation field. Their
detection, with integrated intensities similar to that of the
underlying continuum, suggests then the presence of an irradiating 
UV-field in the region. Indeed, all detected transitions are commonly
found in regions of intense UV-irradiation where the local ISM is
photo-dissociated (\feii, \sii\ plus the three H$_2$ lines; see
Tielens \& Hollenbach~\cite{TH85} and Kaufman et al.~\cite{Kaufman06})
or photo-ionized (\neii, see Morisset et al.~\cite{Mori02}), 
consistent with the spatial distribution of the continuum-subtracted
integrated emission for all detected lines. In all panels of
Fig.~\ref{m160irsmaps} the grey-scale shows, with the exception of 
H$_2$ S(0), that the intensity of the continuum-subtracted integrated 
line fluxes increases away from source~A. We will discuss this in more 
detail below, but we anticipate that this distribution seems to rule out 
any significant role of source~A in the budget of the UV-field in this 
region. 
The H$_2$ rotational lines are more commonly interpreted as
being due to C-shocks, but the extended distribution observed seems
more consistent with a PDR origin. The line ratio analysis using all
the detected lines in Fig. \ref{m160irslines} was carried out using
the PDR Toolbox\footnote{The PDR Toolbox is at
http://dustem.astro.umd.edu/pdrt/index.html} that conveniently
implements a web-based fitting program using the diagnostic diagrams
based on the recent models by Kaufman, Wolfire \& Hollenbach
(\cite{Kaufman06}). Metallicities 2-3 times higher than local ISM
values, and more typical of H{\sc ii} regions and massive star forming
regions in general (Shaver et al. \cite{Shaver83}; Tielens \&
Hollenbach \cite{TH85}) have been used. Most of the line ratios are
consistent with a far-UV field intensity of the order of 100$\leq$
G$_0\leq 1000$, in units of the FUV field intensity in the solar
neighbourhood (Habing~\cite{H68}), for an ISM density in excess of
10$^4$~cm$^{-3}$. The lines which deviate most from defining a
consistent picture are the H$_2$ S(2) and the \feii, which appear to be
a factor of 2-3 stronger than PDR models predictions given the fluxes
of the other lines. The modeled ISM densities are entirely reasonable
for the intra-clump material of a massive star forming region, as
shown from maps in high-density tracing species like HCO$^+$(1-0),
$^{13}$CO(2-1) and CS(3-2) (Brand et al. \cite{Brand01}).

To verify if external field objects could possibly be responsible for
the detected UV-irradiation, we searched the Spectral Classification
catalogues available in Aladin for stars with spectral type earlier
than B. The nearest object is the B9V star BD+60 2600 located more
than 10\amin\ to the North; assuming standard stellar parameters (Thompson
\cite{T84}) and a distance equal to \mol\, the resulting G$_0$ is few
orders of magnitudes below the observed levels.

Among the various detected sources in this region, source~A is the
most massive and luminous; its extreme optical depth below 20\um,
however, would in principle allow only \feii, \sii\ and H$_2$ S(0) to
escape from its inner envelope regions. However, the only indication
we have in this respect, as already noted, is limited to the H$_2$ S(0)
which however traces warm gas in general; the former two lines, that
are the most commonly used PDR tracers, are instead peaking away from
source~A. The fact that also the intensity of the PAH feature, which is
directly related to the intensity of the UV-field, increases away from
source~A confirms that the latter cannot be the radiating source of the UV-field, not
even in the presence of a clumpy circumstellar envelope.

The other sources detected in the region are much brighter than A at
24~\um\ and many have a near-IR counterpart, so that they could be
intermediate mass YSOs in a more evolved state than source~A. However,
given the limited spatial coverage of the IRS maps, it is not possible
to use spectroscopic tools to get more insight into the nature of
these various objects.

More can be learned from the SED modelling of the other sources
revealed at 24\um\ (Table~\ref{fluxtab}). Now, we seek evidence for
the presence of an embedded ZAMS star, as opposed to a pre-MS object,
which might be responsible for the relatively intense UV-irradiation
conditions in the region, as well as for the extended radio continuum
which is thermal free-free in origin (see Fig.~\ref{mipsfig}b). 
We again used the automatic SED fitting tool of Robitaille et al. 
(\cite{Robi07}). For sources with no counterpart in the near-IR or 
in the visible, we adopted the limiting magnitudes of  our own JHK$_s$ 
data (Faustini et al., submitted) and of the DSS2 plates. We searched, 
among the various $\chi^2$-ranked fits, for
models where the central stars have photospheric temperatures and
stellar radii typical of ZAMS stars, to verify if some of the sources
detected in the region could be capable of radiating sufficient
quantities of UV flux to photo-dissociate the intraclump medium as
suggested by the IRS spectroscopy. Such models could be found for
sources B, C, D, H and I, although not with the best formal $\chi^2$; 
they rank lower in the best $\chi^2$ list, but still provide very good
fits (shown in Fig.~\ref{othermodels}). No such fits could instead be
found for the other sources in Table~\ref{fluxtab}, which can only be
fitted with pre-ZAMS objects. Table~\ref{models} reports the model
parameters for SED of embedded ZAMS stars; Col.~9 lists the Lyman
continuum as tabulated by Thompson (\cite{T84}) for ZAMS stars with
photospheric temperatures similar to those reported in Col.~3. Col.~8
reports the model intrinsic bolometric luminosity.  The sum of the
values in Col.~8 plus the luminosity previously estimated for source~A
amounts to $\sim$30\,000\lsun\ which is 50\% higher than the value
obtained for the entire region using the IRAS fluxes; we regard this
discrepancy as barely significant, given the spread of the model
luminosity depending on the exact model selected in the fit. Col.~10
shows the unattenuated integrated 6-13.6~eV far-UV continuum,
expressed in units of the far-UV field intensity in the solar
neighborhood (Habing~\cite{H68}), irradiated by each source at the
position of source~A.

\begin{table}
\setlength{\tabcolsep}{0.04in}
\caption{Models Results for candidate ZAMS Sources in the Field}
\begin{tabular}{lcccccccccc}\hline\hline
Sou. & M$_{\star}$ & T$_{\star}$ & R$_{\star}$ & R$_{Disk}$ & $i$ & A$_V$ & L$_{bol}^a$ & Log[N$_{Ly}$] & G$_0$ \\
 & [M$_{\odot}$] & [K] & [R$_{\odot}$] & [AU] & [\adeg] & [mag] & \lsun &  [s$^{-1}$] &  \\ \hline
B  & 11.4 & 27500 & 4.3 & 380 & 87 & 15.5 & 9400 & 46.3 & 200 \\
C  & 9.7 &  25300 & 3.9 & 2000 & 70 & 11.5 & 5600 & 45.8 & 120 \\
D & 10.0 & 25600 & 4.0 & 660 & 87 & 7.3 & 6400 & 45.9 & 135 \\
H & 6.0 & 18400 & 2.9 & 410 & 87 & 0.0 & 900 & 43.4 & 10 \\
I  & 9.2 & 24500 & 3.8 & 120 & 70 & 6.2 & 4700 & 45.8 & 50 \\ 
\hline
\end{tabular}
\parbox{8cm}{$^a$ As given by the models.}
\label{models}
\end{table}

The total G$_0$ at the position of source~A is consistent with the
regime deduced from the infrared IRS lines, if PDR in origin. Since
G$_0$ in Col.~10 of Table~\ref{models} is computed with no intervening
attenuation between the emitting source and the position of source~A,
we must verify that this is indeed plausible. Indeed, all the models
in Table~\ref{models} are obtained for systems where the ZAMS star is
surrounded by a disk with high inclination angle with respect to the
line-of-sight, and with no envelope. The SEDs which fit the observed
data points (see Fig.~\ref{othermodels}) are obtained assuming an 
additional extinction correction along the line-of-sight to the observer 
(Col.~7) due to intervening dust not related to the immediate circumstellar 
environments. Since these values are not homogeneous, and the distribution 
of optically visible objects in the \mol\ area is not suggestive of 
significant or variable interstellar absorption, we believe this extinction 
to be intra-cluster in origin. It is clear that with
such amounts of extinction the far-UV field should be entirely
absorbed, so that we have to assume that the line-of-sight from the
various sources to source~A must be relatively dust-free. That this
must be the case also seems to be suggested by the distribution of
the radio continuum from thermal free-free emission (dashed lines in
Fig.~\ref{mipsfig}b); the extent of the emission requires that sources
of ionizing continuum must be present, and that this continuum is not
confined to the immediate surroundings of the ionizing stars. The most
likely sources for the radio emission seem to be sources D and H for
the eastern radio lobe, and sources B and C for the western radio lobe.

Concerning the intensity of the radio emission, however, its
conversion into unattenuated Lyman continuum intensity N$_{Ly}$
provides values (Molinari et al. \cite{Moli02}) one order of magnitude
higher than those that can be justified by the modeled ZAMS objects in
the region (Col.~9 of Table~\ref{models}). Even integrating the IRAS 
and millimeter fluxes would yield a value of
$\sim$20\,000~\lsun\ which, also if coming from a single ZAMS star
(which is not the present case), is again not sufficient to provide
the amount of N$_{Ly}$ deduced from the radio fluxes. It should also
be noted that the latter is a lower limit because the free-free emission 
has been assumed optically thin and possible
attenuation from dust inside the HII regions has not been considered. The
simplest way to reconcile this apparent inconsistency would be to
posit that the distance of 4.9~kpc which we assumed for this object is
incorrect. Indeed, while the bolometric luminosity scales with the
square of the distance, the relationship between ZAMS luminosity and
Lyman continuum is much steeper. In the limiting case of a single object
emitting all the luminosity (which is unrealistic given the evidence),
moving the object to a distance of $\sim$6~kpc would raise the
luminosity to $\sim$29\,000~\lsun\ which would produce
Log(N$_{Ly}$)$\sim $47.4; this is equal to the N$_{Ly}$ deduced
from the observed radio flux rescaled to a distance of 6~kpc. The evidence 
from the radio continuum, however, is that there are at least two major 
sources of the ionizing-flux field; a simple calculation shows that in the 
limit of two dominant ZAMS objects the distance should be increased to 
about 8~kpc. The V$_{LSR}$ that would be implied by this distance is more 
than 20~\kms\ different from the value measured from radio spectroscopy, 
and this is not incompatible with the magnitude of the streaming motions 
deduced from the observed velocity field (Brand \& Blitz~\cite{BB93}).

Changing the distance of this region to 8~kpc does not affect our interpretation 
about the nature of source~A as it emerges from Fig.~\ref{lmdiagm160}; both 
envelope mass and luminosity scale the same way with distance so that the 
position of source~A in Fig.~\ref{lmdiagm160} would shift toward upper-right 
along a line of slope 1, therefore leaving the source in the area occupied by 
MM-P objects. This higher distance would also imply that the mass regime of the 
\mol\ region is higher than previously thought; in particular source~A, which 
in our proposed evolutionary interpretation and at the distance of 4.9~kpc 
would reach the ZAMS (the solid line corresponding to the location of the IR-P 
objects in Fig.~\ref{lmdiagm160}) at a luminosity of nearly 16\,000\lsun\ as a 
B0.5 star, if placed at the distance of 8~kpc would reach the ZAMS at about 
40\,000\lsun, roughly corresponding to an O9.5 star.

\section{Conclusions}

Sub-arcminute spatial resolution in the far-IR proves to be critical
for a proper assessment of the evolutionary stage of a massive forming
object. SED modeling of the most prominent YSOs in the \mol\ region
suggests the presence of several intermediate and high-mass YSOs in
different evolutionary stages. While some of them are compatible with
being ZAMS objects with spectral types between B1.5 and B5, source~A
is best interpreted as a strongly accreting object of central mass
comparable to the other ZAMS YSOs in the area, but not yet on
the ZAMS. The region immediately surrounding source~A is bright in
infrared lines which are typical of photo-dissociation/ionisation
regions; the deduced intensity of the radiative far-UV field at the
position of source~A is compatible with emission from the few ZAMS
objects as characterized by the SED modelling, which are also likely
to be responsible for the extended radio continuum emission. There is
an inconsistency, however, between the Lyman continuum estimated from
the radio flux and from the luminosity of the modeled ZAMS stars in the
region. At the moment we can only reconcile this if we assume that the
region is almost twice as distant as assumed up to now, which obviously 
would need independent confirmation.

The extreme values of the circumstellar extinction implied by the
extreme SED of source~A exclude that this source can participate in
any way to the far-UV radiative field which permeates the region. As a
whole, the population of intermediate and massive YSOs in the region
is suggestive of a star formation timescale of the order of few
10$^5$~years, as suggested by models.

However, intermediate- and high-mass stars are not the only objects found
in this region. A cluster of lower-mass stars is revealed in the
near-IR; this is partially apparent in Fig.~\ref{mipsfig}b, where the
contrast, however, is not optimal to visually reveal the much more
abundant population of fainter objects. Stellar density analysis shows
the cluster to be as extended as the extent of the mid- and far-IR
emission, with several tens of members (Faustini et al., 
submitted). The comparison of the K-band luminosity function with an
extensive grid of synthetic cluster models obtained for a wide range
of stellar ages, IMFs and star formation histories, suggests that the
ages of these lower mass objects are at least few 10$^6$~years.

The simultaneous presence of relatively old pre-main sequence objects
with a massive YSO still in an active, pre-ZAMS, accretion phase,
confirms that star formation tends to be an ongoing process for quite
a long time span. The highest-mass star seems to be the last one to
form.

The numerical importance of massive YSOs in a pre-ZAMS phase cannot be
firmly established at the moment. Submillimeter surveys (Hill et
al.~\cite{Hill05}, Beltr\'{a}n et al.~\cite{Beltran06}) reveal a
consistent population of submillimeter cores in the proximity of IRAS
point sources. These cores are devoid of mid-IR emission, but without 
an assessment
of their far-infrared properties which only will allow firm
temperature, luminosity and mass estimates, their nature will remain
elusive. The PACS and SPIRE far-infrared cameras on board the Herschel
satellite will be the ultimate tools to help distinguish between
Hot-Cores, pre-ZAMS objects and quiescent cold cores. The recently
approved Herschel Key-Project Hi-GAL for a complete 60-600\um\
continuum survey of the inner Galactic Plane will be invaluable to
obtain firm statistics of massive YSOs in all evolutionary stages,
thus providing a solid foundation to establish the timeline for the
formation of intermediate- and high-mass stars.

%\begin{acknowledgements}
%\end{acknowledgements}

\end{document}